\documentclass[fleqn,usenatbib]{mnras}

\usepackage{newtxtext,newtxmath}

\usepackage[T1]{fontenc}

\DeclareRobustCommand{\VAN}[3]{#2}
\let\VANthebibliography\thebibliography
\def\thebibliography{\DeclareRobustCommand{\VAN}[3]{##3}\VANthebibliography}

\usepackage{graphicx}	
\usepackage{amsmath}	
\usepackage{threeparttable} 
\usepackage{tabularx} 
\usepackage{color} 
\usepackage{float} 
\usepackage{caption}
\usepackage{url}

\newcommand{\xmm}{\textit{XMM-Newton}\ }        
\newcommand{\narrowhline}[1]{\hline & \\[-#1ex] \ }
\newcommand{\EarthRadii}{R$_\text{E}$}
\newcommand{\EarthMasses}{M$_\text{E}$}
\newcommand{\DensityUnits}{g\,cm$^{-3}$}
\newcommand{\ThisStar}{LTT\,9779}
\newcommand{\ThisPlanet}{LTT\,9779\,b}

\title[LTT 9779 b: surviving in the Neptunian desert]{Survival in the Neptune desert: LTT 9779 b kept its atmosphere thanks to an unusually X-ray faint host star}

\author[J. Fern\'andez Fern\'andez et al.]{
Jorge Fern\'andez Fern\'andez,$^{1,2}$\thanks{E-mail: jorge.fernandez-fernandez@warwick.ac.uk}
Peter J. Wheatley$^{1,2}$\thanks{E-mail: p.j.wheatley@warwick.ac.uk}
George W. King$^{3,1,2}$
James S. Jenkins$^{4,5}$
\\
$^{1}$Department of Physics, University of Warwick, Gibbet Hill Road, Coventry CV4 7AL, UK\\
$^{2}$Centre for Exoplanets and Habitability, University of Warwick, Gibbet Hill Road, Coventry CV4 7AL, UK\\
$^{3}$Department of Astronomy, University of Michigan, Ann Arbor, MI 48109, USA\\
$^{4}$Instituto de Estudios Astrof\'isicos, Facultad de Ingenier\'ia y Ciencias, Universidad Diego Portales, Av. Ej\'ercito 441, Santiago, Chile\\
$^5$Centro de Astrof\'isica y Tecnolog\'ias Afines (CATA), Casilla 36-D, Santiago, Chile\\
}

\date{Accepted XXX. Received YYY; in original form ZZZ}

\pubyear{2023}

\begin{document}
\label{firstpage}
\pagerange{\pageref{firstpage}--\pageref{lastpage}}
\maketitle

\begin{abstract}
The Neptunian desert is a region in period-radius parameter space with very few Neptune-sized planets at short orbital periods. Amongst these, LTT\,9779\,b is the only known Neptune with a period shorter than one day to retain a significant H-He atmosphere. If the Neptune desert is the result of X-ray/EUV-driven photoevaporation, it is surprising that the atmosphere of LTT\,9779\,b survived the intense bombardment of high energy photons from its young host star. However, the star has low measured rotational broadening, which points to the possibility of an anomalously slow spin period and hence a faint X-ray emission history that may have failed to evaporate the planet's atmosphere.  We observed LTT\,9779 with XMM-Newton and measured an upper limit for its X-ray luminosity that is a factor of fifteen lower than expected for its age. We also simulated the evaporation past of LTT\,9779\,b and found that the survival of its atmosphere to the present day is consistent with an unusually faint XUV irradiation history that matches both the X-ray and rotation velocity measurements. We conclude that the anomalously low X-ray irradiation of the one Neptune seen to survive in Neptunian desert supports the interpretation of the desert as primarily a result of photoevaporation. 
\end{abstract}




\section{Introduction}\label{sec:intro}

The Kepler space telescope uncovered thousands of exoplanets during its 9 years of operation, many of them without an analogue in our Solar System.
This included super-Earths and sub-Neptunes in tightly packed systems \citep{Borucki11:kepler-mission}.

These discoveries have revealed a scarcity of Neptune to Saturn sized planets (3--10 \EarthRadii) at short orbital periods of $<3$ days. This region, known as the sub-Jovian or Neptunian desert, has become more statistically significant with further planet discoveries by the K2 and TESS missions \citep[e.g.][]{SazboKiss11:neptune-desert, BenitezLlamblay11:neptune-desert, SanchisOjeda14:neptune-desert, Mazeh16:neptune-desert}.

Since close-in planets are easier to detect in transit surveys, with many detections of hot Jupiters and even Earth-sized worlds with periods under two days \citep{MurrayClay09:hot-jupiters-stable, Mazeh16:neptune-desert}, the Neptunian desert cannot be explained as an observational bias alone, and must be caused by underlying physical mechanisms affecting these planets during their formation and/or evolution.

The origin of the Neptunian desert has been attributed to the inability of planets in this region to hold onto their gaseous envelopes due to atmospheric escape \citep{Kurokawa13:neptune-desert-evap, Kurokawa14:neptune-desert-evap, Ionov18:neptune-desert-evap, OwenLai18:neptune-desert-origin}, a phenomenon that is also thought to give rise to the period-radius valley \citep{LopezFortney12:envelope-model, OwenWu13:radius-valley-evap, Fulton17:radius-valley, OwenWu17:radius-valley-evap}.

Neptune-sized planets that formed in the desert, or migrated there in their youth, would be stripped of their H/He envelopes joining the more abundant population of hot rocky worlds with smaller radii. Hot Jupiters (>10 \EarthRadii), on the other hand, are thought to be stable against atmospheric loss thanks to their deep gravitational potentials \citep{Yelle2004:hot-jupiters-stable, MurrayClay09:hot-jupiters-stable, Vissapragada22:neptune-desert}.
The underlying mechanisms that drive atmospheric escape, however, are still unclear, and several have been proposed.

One such mechanism, photoevaporation, derives the energy for atmospheric escape from the X-ray and extreme ultraviolet radiation (together, XUV) originating in the host star's corona. These high energy photons are readily absorbed by the upper layers of large H/He-rich atmospheres, heating and expanding the gas, and driving a hydrodynamic wind that escapes the planet \citep{Lecavelier07:energy-limited, OwenWu13:radius-valley-evap, LopezFortney14:envelope-model}.

\citet{OwenLai18:neptune-desert-origin} argued that two separate mechanisms are responsible for the formation of the two boundaries of the Neptune desert. Its lower boundary is sculpted by the photoevaporation of the atmospheres of Neptune-sized planets.
The upper boundary of the desert, over which hot Jupiters lie, is the result of a tidal disruption barrier that interrupts the high-eccentricity migration of gas giants.

Alternatively, it has been suggested the Neptunian desert may arise due to the effects of Roche Lobe Overflow (RLO) on hot Neptunes \citep{Jackson17:neptune-desert-rlo, Koskinen22:neptune-desert-rlo}, 
or as a result of high-eccentricity migration \citep{Matsakos16:neptune-migration}.

\begin{figure}
    \centering
    \includegraphics[width=\columnwidth]{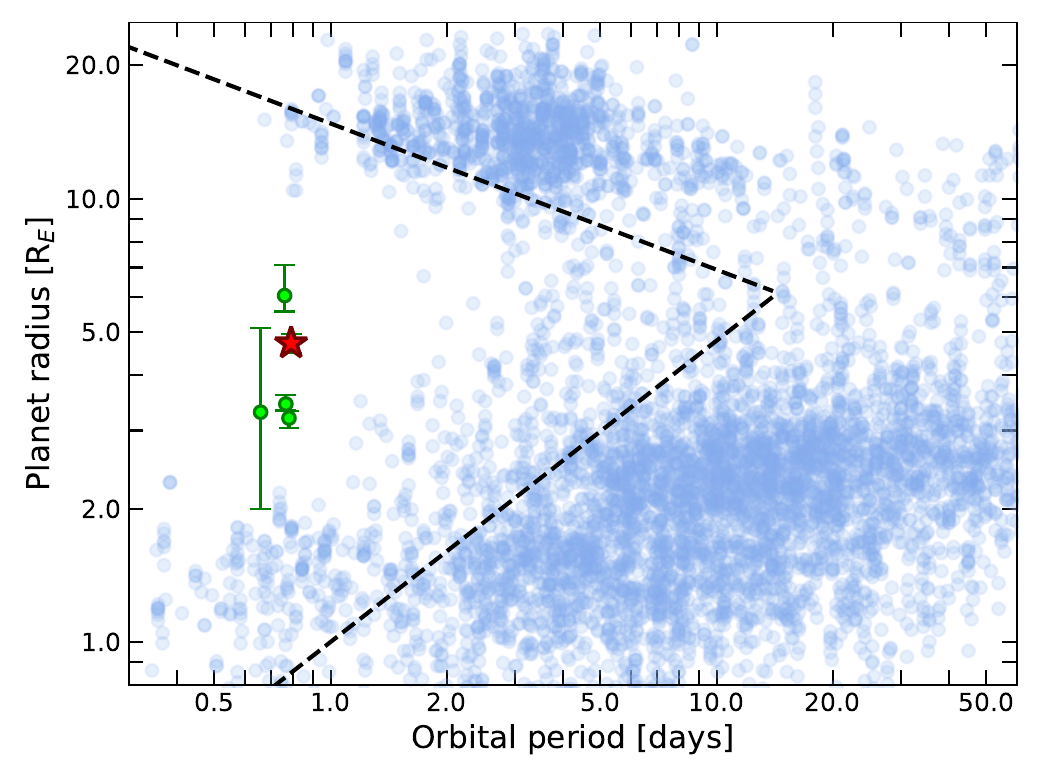}
    \caption[]{Plot of orbital period against planet radius, with \ThisPlanet\ marked as a red star. The upper and lower bounds of the Neptune desert are shown as black dashed lines, following \citet{Mazeh16:neptune-desert}. Data points represent confirmed exoplanets source from the \textit{NASA Exoplanet Archive}, with ultra-short period Neptunes shown as green circles.}
    \label{fig:neptune-desert}
\end{figure}

In Figure \ref{fig:neptune-desert} we plot all validated exoplanets from the NASA Exoplanet Archive\footnote{The NASA Exoplanet Archive can be accessed at \url{https://exoplanetarchive.ipac.caltech.edu/}.} as of July 2023. The Neptune desert is delimited by the boundaries proposed by \citet{Mazeh16:neptune-desert}, who defined the lower and upper edges by finding the boundaries of maximum contrast of planet number density.

In the past few years, a handful of ultra-short period Neptunes have been discovered, a group of planets distinct from the larger exoplanet population, plotted as green points on Figure \ref{fig:neptune-desert}.
Some of these ultra-short period Neptunes
are consistent with a rocky density and hence a high-mass core
(TOI-849\,b, \citet{Armstrong20:TOI-849}; TOI-332\,b, \citet{Osborn2023:toi-332})
and whose origins are still unknown;
others lack mass measurements \citep[K2-399\,b,][]{Christiansen22:K2-399}
or the necessary precision in mass measurement 
to constrain their densities and internal structures \citep[K2-266\,b,][]{Rodriguez18:k2-266b}.
There is also a growing population of low density sub-Neptunes in the desert at periods of 1--3 days,
such as NGTS-4\,b \citep{West19:ngts-4b} and GJ\,1214\,b \citep{Cloutier21:gj-1214b}.

Finally, \ThisPlanet\ (marked in red on Fig.\,\ref{fig:neptune-desert}) is a hot Neptune in an extremely short orbit around a G7V star with a period of just 0.79 days \citep[about 19 hours,][]{Jenkins20:LTT-9779}.
Its measured mass ($29.30^{+0.78}_{-0.81}$\,\EarthMasses) and radius ($4.72\pm0.23$\,\EarthRadii) require a very low density ($1.53\pm0.23$\,\DensityUnits) characteristic of a volatile-rich planet.

The presence of a large gaseous atmosphere has been confirmed by
\citet{Crossfield20:ltt-9779-spitzer} and \citet{Dragomir20:ltt-9779-spitzer} with infrared observations from Spitzer,
\citet{Edwards23:ltt-9779-hst} with HST/WFC3 transmission spectroscopy, and by \citet{Hoyer23:ltt-9779-cheops} with CHEOPS occultation measurements.
The existence of a substantial H/He-rich atmosphere makes \ThisPlanet\ a one-of-a-kind planet and its survival deep in the Neptune desert puzzling.
If the Neptune desert is the result of X-ray driven evaporation, one would expect that a highly irradiated Neptune-sized planet, such as \ThisPlanet, should have already been stripped of its envelope.

We list the stellar parameters of LTT\,9779 in Table\,\ref{tab:star_params} and note that the star has very low measured rotational broadening ($v\sin{i}$). Assuming the rotation axis of the star is not pointing towards the Earth, this implies a spin period of around 45\,d, which is unusually long for a star of its type and age \citep{Johnstone21:rotation-model}. Since stellar activity and X-ray emission are strongly correlated with spin period \citep[e.g][]{Wright11:rotation-xrays} the slow spin of \ThisStar\ suggests diminished XUV emission that might have failed to remove the planet's atmosphere.

In this paper, we test this scenario for the survival of \ThisPlanet. We observed \ThisStar\ with \xmm and placed a limit on its X-ray luminosity (Sect.\,\ref{sec:observations}). We then modelled the rotation and X-ray emission history of the star (Sect.\,\ref{sec:xuv-model}) and the possible evaporation histories of \ThisPlanet\ (Sects.\,\ref{sec:evap-model}\,\&\,\ref{sec:results}). In Sect.\,\ref{sec:discussion}, we conclude that the survival of \ThisPlanet\ in the Neptunian can indeed be explained by an unusually low X-ray emission history of its host star, adding weight to the suggestion that the wider Neptunian desert originates from photoevaporation. 

\begin{table}
    \centering
    \begin{threeparttable}
    \caption{Stellar parameters of \ThisStar.}
    \label{tab:star_params}
    \begin{tabular}{llcc}
        \hline\hline
        Parameter & Units & Value & Reference \\
        \narrowhline{5} \\
        \multicolumn{4}{c}{Astrometric} \\
        RA  (J2000)         & hh:mm:ss      & $23:54:40.21$      & Gaia DR3\tnote{c} \\
        Dec (J2000)         & dd:mm:ss      & $-37:37:40.52$     & Gaia DR3\tnote{c} \\
        $\mu_{\text{RA}}$   & mas yr$^{-1}$ & $247.634\pm0.013$  & Gaia DR3\tnote{c} \\
        $\mu_{\text{Dec}}$  & mas yr$^{-1}$ & $-69.752\pm0.014$  & Gaia DR3\tnote{c} \\
        Parallax            & mas           & $12.338\pm0.017$   & Gaia DR3\tnote{c} \\
        Distance            & pc            & $81.050\pm0.011$   & Gaia DR3\tnote{c} \\
        \narrowhline{5} \\
        \multicolumn{4}{c}{Physical} \\
        Spectral type    & ---         & G7V                    & J20\tnote{a} \\
        V                & mag         & $9.76 \pm 0.03$        & UCAC4\tnote{b} \\
        T$_{\text{eff}}$ & K           & $5480 \pm 42$          & J20\tnote{a} \\
        M$_*$            & M$_{\odot}$ & $1.02^{+0.02}_{-0.03}$ & J20\tnote{a} \\
        R$_*$            & R$_{\odot}$ & $0.949 \pm 0.006$      & J20\tnote{a} \\
        L$_*$            & L$_{\odot}$ & $0.708 \pm 0.016$      & J20\tnote{a} \\
        {[Fe/H]}         & dex         & $0.25 \pm 0.04$        & J20\tnote{a} \\
        $v\sin{i}$       & km s$^{-1}$ & $1.06\pm0.37$          & J20\tnote{a} \\
        P$_{\text{rot}}$ & days        & $<45$                  & J20\tnote{a} \\
        Age              & Gyr         & $2.0 ^{+1.3} _{-0.9}$  & J20\tnote{a} \\
        $\log{\text{R}'}_\text{HK}$    & dex  & $-5.10\pm0.04$  & J20\tnote{a} \\
        \hline
    \end{tabular}
    \begin{tablenotes}
        \item[a] J20: \citet{Jenkins20:LTT-9779}
        \item[b] UCAC4: \citet{simbad-ucac4}
        \item[c] Gaia DR3: \citet{gaia-dr3}
        \item[d] 2MASS: \citet{2mass}
    \end{tablenotes}
    \end{threeparttable}
\end{table}

\begin{table*}
    \centering
    \captionsetup{justification=centering}
    \begin{threeparttable}
    \caption{Upper limit count rates for the PN and MOS1+2 instruments.}
    \label{tab:upper-limits}
    \begin{tabular}{l c c c c c c}
        \hline \hline
        Instrument & Energy range & Exposure & Source & Expected background
        & 90\% upper limit & Upper limit rate \\
         & keV & ks & counts & counts
        & source counts & counts ks$^{-1}$ \\
        \hline
        pn     & 0.15 -- 2.0 & 31.7 & 91 & 69.6
               & 37.3 & $1.177$ \\
        MOS1+2 & 0.20 -- 2.0 & 42.3 & 32 & 34.3
               & 9.64 & $0.228$ \\
        \hline
    \end{tabular}
    \footnotesize
    \end{threeparttable}
\end{table*}

\begin{table}
    \centering
    \begin{threeparttable}
    \caption{Upper limit X-ray fluxes and luminosities predicted by the APEC model from pn and MOS count rates.}
    \label{tab:xray-fluxes}
    \begin{tabular}{c l c c}
        \hline \hline
        Energy range & Instrument & X-ray flux & Luminosity \\
        (keV) &  & ($10^{-15}$ erg cm$^{-2}$ s$^{-1}$) & ($10^{27}$ erg s$^{-1}$) \\
        \hline
        0.1 -- 2.4 & pn     & 2.87 & 2.26 \\
                   & MOS1+2 & 2.83 & 2.23 \\
        \hline
        0.2 -- 2.4 & pn     & 2.07 & 1.63 \\
                   & MOS1+2 & 2.04 & 1.60 \\
        \hline
    \end{tabular}
    \end{threeparttable}
\end{table}

\section{X-ray observations \& analysis}\label{sec:observations}

\ThisStar\ was observed with the \xmm spacecraft on May 27th and 28th 2021 for a continuous exposure of 52\,ks (ObsID: 0884250101, PI: Wheatley).
The star, however, was not detected in the standard pipeline outputs.
We inspected the data from the the EPIC pn \citep{epic-pn} and MOS \citep{epic-mos} detectors using the proper-motion corrected coordinates of the star on the pipeline-processed event list, adopting both wide and narrow standard energy bands. Nevertheless, we found no evidence of a detection of our target. 

We also found that the particle background in the detectors increased significantly towards the end of the observation, and hence we decided to remove these noisy sections by dropping the time intervals with count rates higher than $0.5\rm\,s^{-1}$ for pn and $0.35\rm\,s^{-1}$ for MOS in the band 10-12\,keV across the full area of the detector. The remaining data amounted to 31.7\,ks for pn and 42.3\,ks for each MOS. The target star, however, remained undetected after this removal of noisy data.

In the absence of a detection, we estimated an upper limit X-ray flux from \ThisStar\ using the method described by \citet{Farihi18:xray-upper-limit}.
This analysis is also based on the standard data analysis threads using the XMM Science Analysis Software (SAS)\footnote{Users Guide to the XMM-Newton Science Analysis System", Issue 17.0, 2022 (ESA: XMM-Newton SOC), \url{https://www.cosmos.esa.int/web/xmm-newton/sas-threads}} version 19.

We first applied an extraction area around the proper motion corrected coordinates of the source with a circular aperture of radius 16 arcseconds, as well as a secondary aperture of radius 80 arcseconds for the background in a nearby region that contained no X-ray sources. We collected the detected counts in these regions using the energy ranges 0.15 -- 2.0\,keV for pn and 0.2 -- 2.0\,keV for MOS.
We discarded counts over 2\,keV as coronal emission from stellar activity is relatively soft \citep{Gudel04:xray-temp-relation}. 
We then scaled the background counts to the aperture area used for the source region. In the case of the two MOS detectors, we combined the count rates from both instruments by making use of common Good Time Intervals (GTI) and adding up the detected counts.

We then applied the method by \citet{Kraft91:low-counts} to estimate 90\% confidence upper limit count rates. They model noise using a Poisson distribution and adopt a Bayesian approach with the prior that the source count rate cannot be negative. The results, shown in Table \ref{tab:upper-limits}, indicate upper limit count rates of $1.18$\,ks$^{-1}$ for pn and $0.23$\,ks$^{-1}$ for the combined MOS1 and MOS2 detectors (hereafter MOS1+2).

In order to convert the count rates to X-ray fluxes, we built model spectra using the software \texttt{XSPEC} version 12.12.1 \citep{xspec}, which we then scaled to match the upper limit count rates in Table\,\ref{tab:upper-limits}.
In the case of MOS, we loaded the response and ancillary files for MOS1 only, and scaled the model to half the combined MOS1+2 count rate.
We adopted solar abundances from \citet{Asplund09:solar-abund} and accounted for interstellar absorption with the TBABS model by \citet{tbabs}. We estimated an interstellar hydrogen column density of $2.5\times10^{19}$\,cm$^{-2}$ following \citet{RedfieldLinksy01:ism-hydrogen}, who determined a hydrogen density of 0.1\,cm$^{-3}$ for the local interstellar medium.
We used a single-temperature APEC model to simulate the spectrum \citep{apec}, which models the emission spectrum of collisionally-ionized diffuse gas.
In order to determine a plasma temperature for our spectral model, we adopted the luminosity-temperature relation by \citet{Gudel04:xray-temp-relation}, which shows that average coronal temperatures increase with X-ray luminosity on Sun-like stars.
We found that their relation is consistent with our choice of model for plasma temperatures between 0.15 and 0.3\,keV.
For our upper limit calculation we selected a temperature of 0.15\,keV, since this results in the most conservative (higher) limit. 

Finally, we scaled and integrated the spectral model, and estimated X-ray fluxes and luminosities for both the pn and MOS1+2 instruments, as shown on Table \ref{tab:xray-fluxes}. The pn and MOS limits are very similar, corresponding to upper limit X-ray luminosities of $2.2\times10^{27}$\,erg s$^{-1}$ in the 0.1--2.4\,keV energy range, and $1.6\times10^{27}$\,erg s$^{-1}$ in the 0.2--2.4\,keV band.

The measured rotational velocity $v\sin{i}$ of \ThisStar\ implies a spin period of 45\,d, assuming its spin axis is aligned with the planet's orbit.
Using rotation-activity relations \citep[e.g.][]{Wright11:rotation-xrays}, we estimated an X-ray luminosity of $3.9\times10^{27}$\,erg s$^{-1}$ in the band 0.1--2.4\,keV, which is within a factor of two of our upper limit -- consistent with the $1\sigma$ uncertainty on the measured $v\sin{i}$.

We also verified that the exact value for the hydrogen column density did not have a strong effect on the spectral model for values of $N_H\leq10^{21}$\,cm$^{-2}$. A change of a factor of two in this quantity resulted in only a difference of a few percent in the output fluxes.

\begin{figure*}
    \includegraphics[width=\textwidth]{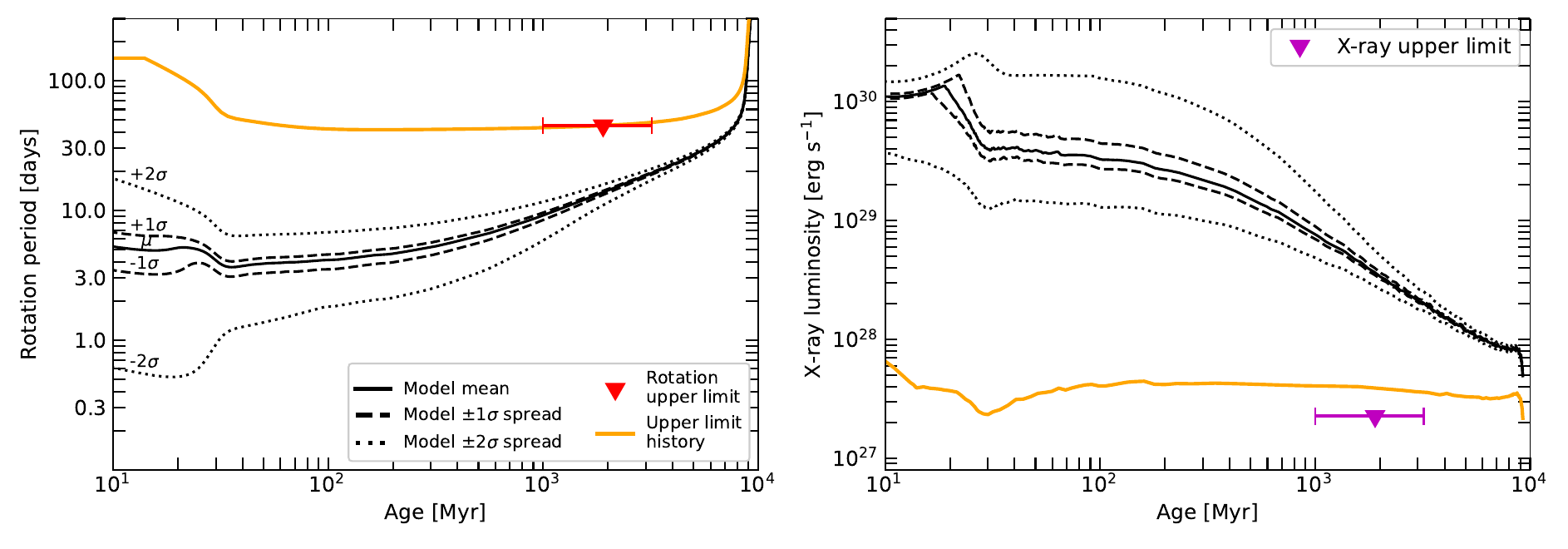}
    \caption{
    \textbf{Left panel}: spin histories for a 1 M$_\odot$ star predicted by \citet{Johnstone21:rotation-model}, including the population mean, $1\sigma$, and $2\sigma$ spreads, as well as the spin history fitted to the 45-day upper limit spin of \ThisStar.
    \textbf{Right panel}: X-ray emission histories in the 0.1--2.4\,keV band corresponding to the spin histories on the left panel. Our X-ray upper limit measured from \xmm observations is shown as a magenta triangle.}
    \label{fig:stellar-evo}
\end{figure*}

\section{Stellar rotation and XUV history}\label{sec:xuv-model}

In order to model the XUV emission history of \ThisStar, we adopted the rotation evolution models by \citet{Johnstone21:rotation-model}.
Those authors model the spin evolution of stars of different masses by combining angular momentum transfer mechanisms within the star (core-envelope coupling) as well as interactions with its environment (stellar wind and disc-locking),
and constrain the distribution of stellar rotation rates to observations of clusters of ages 12 Myr to 2.5 Gyr.
Their models predict that the average rotation period for Solar-mass star of 2\,Gyr age, such as \ThisStar, is  $14$\,days, with a $2\sigma$ spread from $12$ to $16$\,days. The spin and X-ray luminosity evolution of a $1\text{M}_\odot$ star according to this model are shown in black in Figure \ref{fig:stellar-evo}, with the the X-ray emission determined from the spin evolution using rotation-activity relations \citep{Pizzolato03:xray-rotation, Wright11:rotation-xrays, Wright18:rotation-xrays}.

Even though the rotation period of \ThisStar\ is unknown, \citet{Jenkins20:LTT-9779} were able to place an upper limit on the period of 45\,days based on a HARPS $v\sin{i}$ of $1.06\pm0.37$\,km s$^{-1}$.
This 45\,d period will correspond to the true rotation period of the star if its spin axis is aligned with the orbit of the planet (i.e. $i=90^{\circ}$). The implied period remains well above 14\,d for stellar inclinations greater than 20\,deg, and the probability of an inclination of 20\,deg or less is only 6\%, even if the stellar inclination is randomly misaligned with the planet's orbit.
Furthermore, \citet{Jenkins20:LTT-9779} also measured the strength of the \ion{Ca}{II} HK lines and found a $\log{R'}_{HK}$ value of $-5.10\pm0.04$\,dex, which is indicative of a very low activity and also supports the star having an unusually slow rotation period.

In order to model a plausible low XUV emission history, we choose to adopt the 45-day limit as the current rotation period of the star and use the models by \citet{Johnstone21:rotation-model} to extrapolate to earlier and later times. 
The spin and X-ray luminosity evolution in this scenario are shown in orange in 
Figure \ref{fig:stellar-evo}.

Our results for a typical star in Figure \ref{fig:stellar-evo} show a predicted X-ray luminosity of $3.4\times10^{28}$ erg s$^{-1}$ 
for \ThisStar\ at the present time, which is a factor of 15 brighter than our \xmm\ upper limit from Sect.\,\ref{sec:observations}. 
In contrast, our X-ray upper limit is within a factor of 2 of the predicted X-ray luminosity for the 45\,d spin period corresponding to the measured $v\sin{i}$ and an aligned system. 
Our \xmm\ upper limit is clearly more consistent with the anomalously slow rotation scenario.

\begin{figure}
    \centering
    \includegraphics[width=\columnwidth]{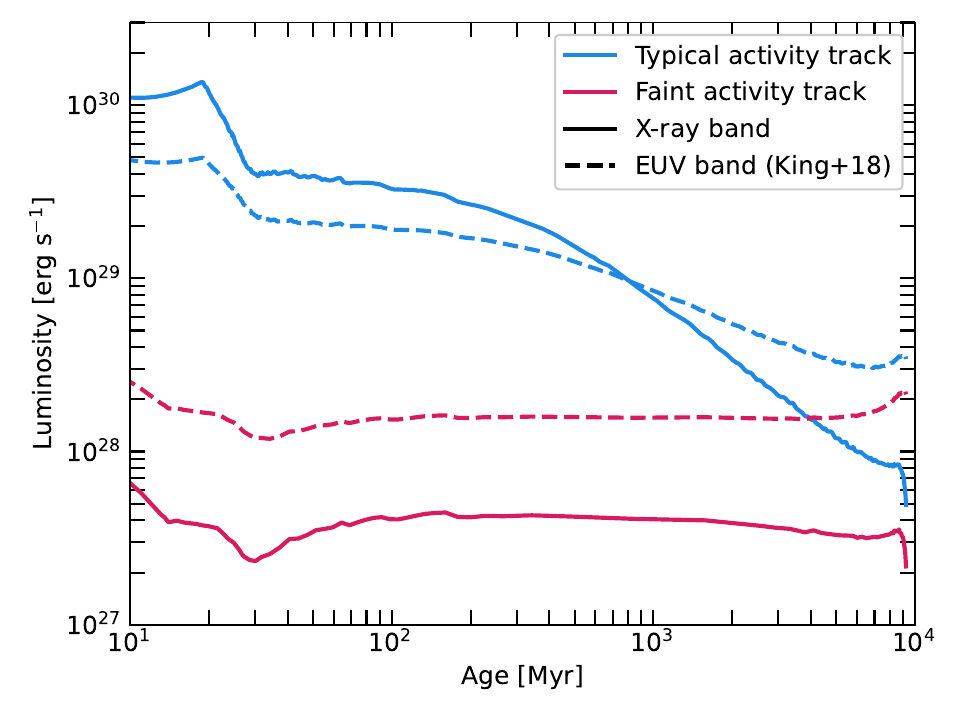}
    \caption[]{Emission histories for the X-ray band (solid lines) and EUV band (dashed lines) for the two stellar models discussed in Section~\ref{sec:xuv-model}: the predicted history for a solar mass star (blue) and the fainter history motivated by spin and X-ray measurements (red).
    The EUV band was estimated from the X-rays using the scaling law by \citet{King18:euv-relation}.}
    \label{fig:stellar-euv}
\end{figure}

\section{Evolution modelling}\label{sec:evap-model}

We used the X-ray luminosity evolution for the different scenarios shown in Figure~\ref{fig:stellar-evo} to model the evaporation history of the \ThisPlanet\ in order to determine the conditions under which the planetary
atmosphere will be stripped.

We used the \texttt{photoevolver} \footnote{The code is available on GitHub at \url{https://github.com/jorgefz/photoevolver}} code \citep{Fernandez23},
which combines three ingredients:
(1) a description of the stellar X-ray history, using the model by \citet{Johnstone21:rotation-model} described in Section \ref{sec:xuv-model},
(2) the envelope structure model by \citet[][Eqn.\,5]{ChenRogers16:envelope-model} based on MESA hydrodynamic simulations, which links the atmospheric mass to its size, and
(3) the mass loss model by \citet{Kubyshkina18:mass-loss-model}, which translates incident X-ray irradiation into atmospheric mass loss rate.

At each simulation time step, the XUV flux incident on the planet is drawn from the stellar tracks in Figure~\ref{fig:stellar-evo} and then used to remove some mass from the atmosphere, calculated using the mass loss model. The envelope size is then recalculated with the new mass, and the age advances one time step forward.
We evolved the planet from the age of 10 Myr, the time at which the protoplanetary disc has fully dissipated \citep{Fedele10:disk-dispersal-10myr} and any boil-off processes have completed \citep{Lammer16:boil-off, OwenWu16:boil-off}, to 5 Gyr, with a fixed time step of 0.1 Myr. We considered the envelope stripped when its mass reached a value below 0.01\% of the planet's mass, as it is the lower limit for which the envelope model by \citet{ChenRogers16:envelope-model} is rated.

We adopted two spin period histories from Figure~\ref{fig:stellar-evo} to find feasible evaporation pasts for \ThisPlanet: one matching the typical spin period distribution for a solar-mass star, and another matching the star's measured rotation upper limit of 45 days at its current age.
We also estimated the extreme ultraviolet (EUV) stellar emission in the band 0.0136--0.1\,keV using the empirical relations by \citet{King18:euv-relation}, which are based on Solar \textit{TIMED/SEE} data.
We plot the relative contributions of the X-ray and EUV bands on Figure~\ref{fig:stellar-euv}.

\begin{table}
    \centering
    \begin{threeparttable}
    \caption{Planetary parameters and internal structure of \ThisPlanet.}
    \label{tab:structure}
    \addtolength{\tabcolsep}{-2pt}
    \begin{tabular}{l l c c}
        \hline \hline
        \multicolumn{2}{c}{Parameter} & Value & Source \\
        \hline
        Orbital period    & P$_\text{orb}$ (days)           & $0.792\pm(9.3\times10^{-6})$ & J20\tnote{a} \\
        Semi-major axis   & a (AU)                          & $0.01679\pm0.00014$          & J20\tnote{a} \\
        Equilibrium temperature & T$_\text{eq}$ (K)         & $1978\pm19$                  & J20\tnote{a} \\
        Mass              & M$_\text{p}$ (\EarthMasses)     & $29.30^{+0.78}_{-0.81}$      & J20\tnote{a} \\
        Radius            & R$_\text{p}$ (\EarthRadii)      & $4.72\pm0.23$                & J20\tnote{a} \\
        \hline
        Core radius       & R$_\text{core}$ (\EarthRadii)   & $2.69\pm0.11$   & This work \\
        Core mass         & M$_\text{core}$ (\EarthMasses)  & $27.30\pm1.10$  & This work \\
        Envelope radius   & R$_\text{env}$  (\EarthRadii)   & $2.03\pm0.26$   & This work \\
        Envelope mass fraction & f$_\text{env}$             & $0.067\pm0.029$ & This work \\
        \hline
    \end{tabular}
    \addtolength{\tabcolsep}{2pt}
    \footnotesize
    \begin{tablenotes}
        \item[a] J20: \citet{Jenkins20:LTT-9779}
    \end{tablenotes}
    \end{threeparttable}
\end{table}

\begin{figure*}
    \includegraphics[width=\textwidth]{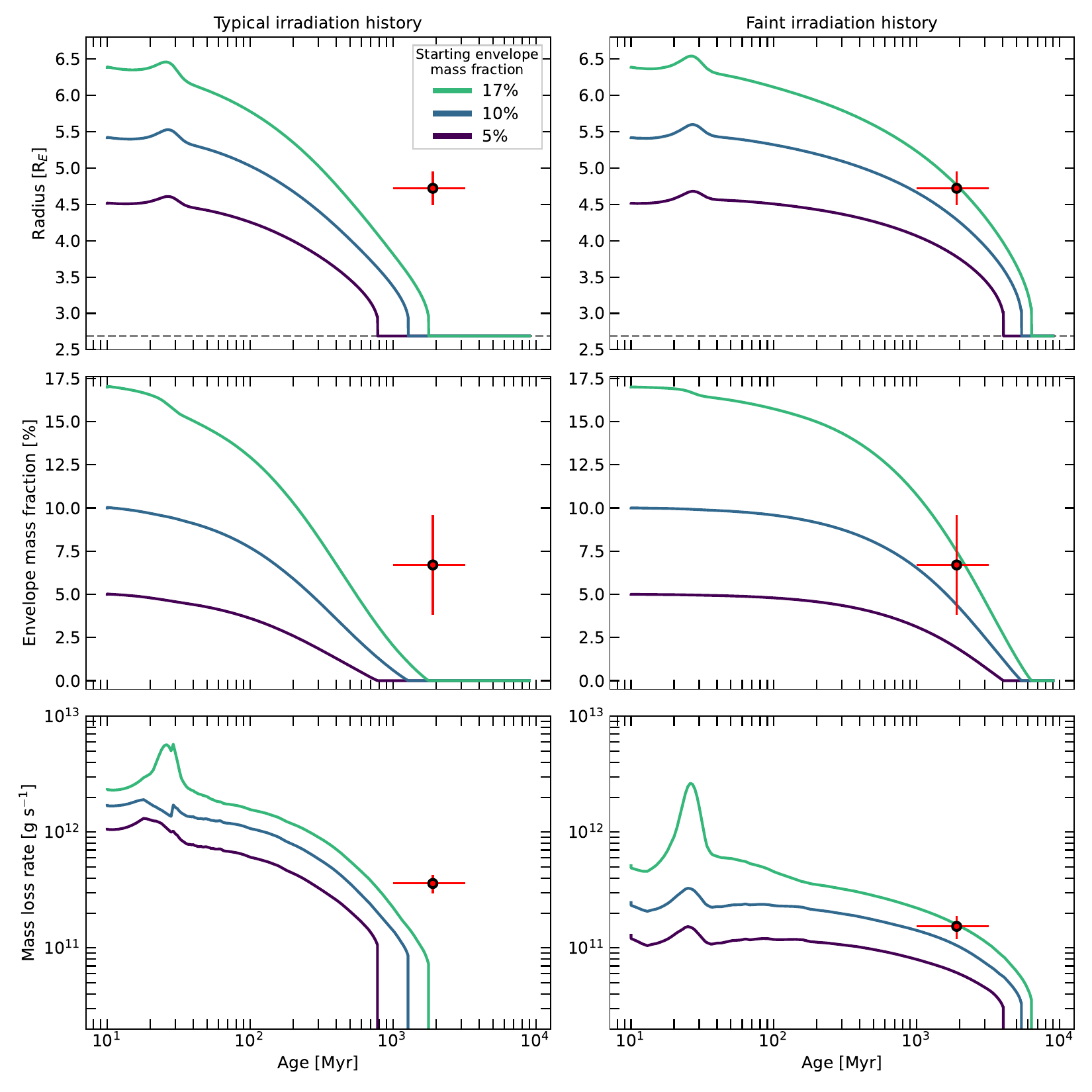}
    \caption{
    Evaporation histories of a range of simulated scenarios for \ThisPlanet, with initial envelope mass fractions from 5\% to 17\%.
    The panels present the evolution of the planet radius (top), envelope mass fraction (middle), and mass loss rate (bottom) under two XUV irradiation histories: (1) the expected stellar emission history of a solar mass star (left column), and (2) a fainter emission history consistent with the measured stellar rotation velocity and our X-ray upper limit from our \xmm observation (right column).
    The dashed grey line on the top panels represents the rocky core radius of \ThisPlanet.}
    \label{fig:planet-evo}
\end{figure*}

\subsection{Internal structure}\label{sec:internal-structure}

We modelled the internal structure of \ThisPlanet\ assuming a silicate-iron (rocky) core surrounded by a large H/He-rich atmosphere, following \citet{RogersOwen21:radius-valley-origin}. This description entails a total of four model parameters: the core radius R$_\text{core}$, core mass M$_\text{core}$, envelope radius R$_\text{env}$, and envelope mass fraction $\text{f}_\text{env}$ (see Table \ref{tab:structure}).
With knowledge of the planet's mass and radius alone (together with orbital and host star parameters), we can deduce values for these quantities by combining four equations. First we defined planet radius as $\text{R}_\text{p}=\text{R}_\text{core}+\text{R}_\text{env}$, and envelope mass fraction as the ratio of envelope and planet masses $\text{f}_\text{env} = \text{M}_\text{env}/\text{M}_\text{p} = (\text{M}_\text{p} - \text{M}_\text{core})/\text{M}_\text{p}$. 
Additionally, we adopted the empirical mass-radius relation for rocky planets by \citet[][Eqn.\,1]{Otegi20:mass-radius} to link the core mass $\text{M}_\text{core}$ to the core radius $\text{R}_\text{core}$. Finally, we adopt the envelope structure model by \citet[][Eqn.\,5]{ChenRogers16:envelope-model}, which links the mass fraction of the envelope $\text{f}_\text{env}$ to its thickness $\text{R}_\text{env}$.
We found that \ThisPlanet\ is consistent with a rocky core of radius $2.69\pm0.11$\ \EarthRadii\, and a gaseous envelope of mass fraction $6.7\pm2.9\%$ (see Table \ref{tab:structure}), in good agreement with the mass fraction of 9\% estimated by \citet{Jenkins20:LTT-9779}. 

Moreover, we estimated that the predicted XUV emission for \ThisStar\ induces a present-day mass loss rate of $3.5\times10^{11}$ g s$^{-1}$ using the model by \citet{Kubyshkina18:mass-loss-model}. Energy-limited photoevaporation, a simpler model that accounts only for the energy from incident X-ray photons \citep{Watson81:energy-limited, Lecavelier07:energy-limited}, predicts a lower present-day mass loss rate of $2.8\times10^{11}$ g s$^{-1}$ using the same evaporation efficiency as \citet{Kubyshkina18:mass-loss-model} of 15\%.
On the other hand, the fainter XUV emission motivated by our \xmm\ observations yields a present-day mass loss rate of $1.5\times10^{11}$ g s$^{-1}$ using the model by \citet{Kubyshkina18:mass-loss-model}, and $0.6\times10^{11}$ g s$^{-1}$ with the energy-limited formulation.

\section {Evaporation history}\label{sec:results}

We simulated a range of evolutionary scenarios for \ThisPlanet, all using the rocky core determined in Table \ref{tab:structure}, but each starting off at an age of 10\,Myr with a different envelope mass fraction.  
We used envelope mass fractions of 5\%, 10\%, and 17\%, with the 
lowest of these values being motivated by the uncertainty on the planet's current envelope mass fraction.

The maximum initial envelope is motivated by the maximum radius the planet could have following its formation.
This is set by the radius of its Roche lobe which we calculate using the expression by \citet[][Eqn.\,3]{OwenLai18:neptune-desert-origin} for low density gaseous planets.
This gives a maximum envelope mass fraction of 17\% for \ThisPlanet\ at the beginning of our simulation.
Finally, the choice of initial envelope mass fraction of 10\% serves as a mid-point between the two values above.

\subsection{Expected XUV history}
\label{sec:expected} 

Firstly, we adopt the XUV luminosity tracks that are expected for a typical solar-mass star following the model by \citet{Johnstone21:rotation-model}, as shown in Figure \ref{fig:stellar-evo} (right hand panel), and evolve \ThisPlanet\ under photoevaporation for the three initial envelope mass fractions.
The results, shown in Figure \ref{fig:planet-evo} (left column)
show that none of these scenarios match the mass and radius of \ThisPlanet\ 
at the present time,
as they are all stripped of their envelopes before now.
This confirms that a planet of this nature orbiting \ThisStar\ with the expected X-ray emission history would already be a hot rocky world of radius $2.7$\,\EarthRadii, joining the population of planets that define the lower edge of the Neptune desert.
We thus rule out this scenario as a description of the evaporation history of \ThisPlanet.

\subsection{Faint XUV history}
\label{sec:faint}

We repeat the analysis in Sect.\,\ref{sec:expected} but adopting a stellar XUV history that respects both the measured stellar rotation velocity and our X-ray upper limit, as determined with our \xmm observation (Sect.\,\ref{sec:observations}). This emission history, shown in Figure \ref{fig:stellar-evo} (right hand panel), matches a star that has a rotation period of 45 days at the age of 2\,Gyr.

We find that the planetary atmosphere survives evaporation to the present day for all three initial envelope fractions, as shown in Figure \ref{fig:planet-evo} (right column). Furthermore, we find that, under these conditions, \ThisPlanet\ is consistent with an evaporation history where it started out with an envelope mass fraction of 10\% to 17\%. This would make the planet a 6--7\,\EarthRadii\,puffy super-Neptune after disc dispersal.

We find that continued exposure to these low XUV fluxes does end up stripping the planet of its envelope by the age of 6--7\,Gyr (Figure~\ref{fig:planet-evo}).
As can be seen in Figure~\ref{fig:stellar-evo}, this is because the low activity track is flat in comparison to the high activity track, and so the X-ray emission in both histories become comparable at later ages.
This behaviour can be attributed to faster rotators spinning down more rapidly than slow rotators, 
leading to the tracks converging \citep{Johnstone21:rotation-model}.

In principle, our conclusion that the low activity track is consistent with the planet retaining its envelope to the present day is sensitive to our choice of EUV scaling law. This is illustrated in Fig.~\ref{fig:stellar-euv}, where it is clear that the EUV emission is expected to be higher than X-rays throughout the lifetime of the star. \citet{Maggio23:V1298-xuv} compared different X-ray/EUV scaling laws and confirmed the that the \citet{King18:euv-relation} relation used here is in good agreement with available EUV observations. Some other EUV scaling laws, however, such as that from \citet{SanzForcada11:euv-relation}
predict somewhat higher EUV emission up to a factor 2 \citep[revised down to 1.5 by ][]{SanzForcada22:euv-law}. In order to test whether the choice of scaling relation affects our conclusions, we re-ran our model with both the \citet{SanzForcada11:euv-relation} and \citet{SanzForcada22:euv-law} relations. We found that the gaseous envelope of the planet is still expected survive to the present day for both scaling relations, demonstrating that our results in practice are not sensitive to this choice.

\section{Discussion \& Conclusions}\label{sec:discussion}

We have presented an analysis of X-ray driven photoevaporation of the ultra-short period Neptune \ThisPlanet: the only known planet deep in the Neptune desert with a significant gaseous envelope.
Since the Neptune desert is thought to be cleared out by X-ray photoevaporation \citep[e.g.][]{OwenLai18:neptune-desert-origin}, the existence of this planet at such a short period is puzzling. 

Specifically, we have considered the possibility that \ThisPlanet\ has survived at its present orbital separation due to anomalously low X-ray emission from its host star, as suggested by its unusually low rotational velocity (Table~\ref{tab:star_params}) and lack of \ion{Ca}{II} HK emission lines \citep{Jenkins20:LTT-9779}.

We made an \xmm observation of the host star and measured an upper-limit to the X-ray luminosity that is a factor of fifteen lower than the expected X-ray luminosity for a star of this type and age (Sects.\,\ref{sec:observations}\,\&\,\ref{sec:xuv-model}). In contrast, our X-ray upper limit is within a factor of 2 of the luminosity expected for a star rotating with the much slower 45\,d period suggested by the rotational velocity and an assumed alignment between the stellar spin and planetary orbital axes. 

We also simulated the possible evaporation history of the planet using a range of initial envelope mass fractions. 
As expected, we confirm that the planetary atmosphere would not survive to the present day under the expected X-ray emission history of a solar mass star like \ThisStar\ (Sect.\,\ref{sec:expected}).
However, we find that a dimmer X-ray past, motivated by the upper limits on the X-ray luminosity and spin period, allows the planetary atmosphere to survive to the present day (Sect.\,\ref{sec:faint}).

We conclude that \ThisStar\ most-likely formed as an anomalously slowly rotating star, and that its close-in Neptune-sized planet \ThisPlanet\ was thus able to survive in the Neptune desert to the present day due to unusually low X-ray irradiation. 
This scenario is consistent with the planet forming in situ and/or migrating within the protoplanetary disc, offering an alternative
to the scenario suggested by \citet{Jenkins20:LTT-9779}, in which late inward migration followed by Roche-lobe overflow (RLO) may have eroded a Jupiter-mass planet down to a hot Neptune like \ThisPlanet.

Moreover, the super-solar ($400\times$) metallicity of its atmosphere together with its very high albedo, determined by \citet{Hoyer23:ltt-9779-cheops}, would make the planet more resistant to evaporation as heavier species require more energy to remove \citep{Wilson22:toi-1064, OwenJackson12:xray-evap}.

Even though the implied slow 45-day spin of \ThisStar\ may be unusual,
examples of long-period (30-45\,d) solar-mass stars have been found in the 950\,Myr-old cluster NGC 6811, hinting towards a rare mechanism that yields extremely slow rotators
\citep{GodoyRivera21:cluster-rotations}.

This unusual feature of the host star is consistent with our interpretation because of the unique nature of \ThisPlanet\ within the Neptunian desert. The strong selection bias in favour of large close-in transiting exoplanets means that planets like \ThisPlanet\ must be extremely rare. In contrast, finding similarly anomalous stars without a hot Neptune is extremely difficult. This is because their low stellar activity results in low amplitude photometric modulation that is difficult to detect. Their long periods are also beyond the reach of most photometric surveys, and we cannot use $v \sin i$ because of contamination by pole-on faster rotators. Nevertheless, it is known that stars are formed with a wide range of initial spin periods, at least in part due to interactions with long-lived protoplanetary discs ($\sim$10\,Myr), which can prevent them from spinning-up during the pre-main sequence stage of their evolution and allowing slow rotators to maintain their initial spin into the main sequence \citep[e.g.][]{Moraux13:disk-locking-bimodal, Ribas15:disk-locking, Richert18:disk-locking}.

Finally, our conclusion that the only known planet deep in the Neptunian desert with a gasesous envelope is also unusual in having an X-ray faint star, strongly supports the suggestion that the primary origin of the Neptunian desert is X-ray driven photoeveporation.

\section*{Acknowledgements}

The work presented here was supported by the UK Science and Technology Facilities (STFC).
JFF is supported by the STFC studentship grant ST/W507908/1. PW acknowledges support under the STFC consolidated grants ST/T000406/1 and ST/X001121/1.
JSJ gratefully acknowledges support by FONDECYT grant 1201371 and from the ANID BASAL project FB210003.
This research made use of the Python packages \texttt{\,numpy} \citep{numpy}, \texttt{\,astropy} \citep{astropy}, \texttt{\,scipy} \citep{scipy}, \texttt{\,matplotlib} \citep{matplotlib}, \texttt{\,Mors} \citep{Johnstone21:rotation-model}, and \texttt{uncertainties}\footnote{Uncertainties: a Python package for calculations with uncertainties, Eric O. LEBIGOT, \url{http://pythonhosted.org/uncertainties}}.
This research has made use of the NASA Exoplanet Archive, which is operated by the California Institute of Technology, under contract with the National Aeronautics and Space Administration under the Exoplanet Exploration Program.
Based on observations obtained with XMM-Newton, an ESA science mission with instruments and contributions directly funded by ESA Member States and NASA.
We thank the reviewer for their constructive comments which have improved the quality of this work.

\section*{Data Availability}


The \xmm data used in this work is publicly available at the \xmm Science Archive (XSA) \footnote{\url{https://www.cosmos.esa.int/web/xmm-newton/xsa}}, under the observation ID 0884250101 (target: \ThisStar, PI: Wheatley).



\bibliographystyle{mnras}
\bibliography{main} 




\appendix



\bsp	
\label{lastpage}
\end{document}